\documentclass[iop]{emulateapj}
\usepackage{amsmath}
\usepackage{ulem}

\begin{document}

\title{Bayesian techniques for comparing time-dependent GRMHD simulations to variable Event Horizon Telescope observations}

\author{Junhan Kim\altaffilmark{1}, 
Daniel P. Marrone\altaffilmark{1}, 
Chi-Kwan Chan\altaffilmark{1}, 
Lia Medeiros\altaffilmark{1,2}, 
Feryal {\"O}zel\altaffilmark{1},
and Dimitrios Psaltis\altaffilmark{1}}
\email{E-mail: junhankim@email.arizona.edu}

\altaffiltext{1}{Department of Astronomy and Steward Observatory, University of Arizona, 933 N. Cherry Ave., Tucson, AZ 85721, USA}
\altaffiltext{2}{Department of Physics, Broida Hall, University of California Santa Barbara, Santa Barbara, CA 93106, USA}

\begin{abstract}
The Event Horizon Telescope (EHT) is a millimeter-wavelength,
very-long-baseline interferometry (VLBI) experiment that is capable of observing
black holes with horizon-scale resolution. Early observations have
revealed variable horizon-scale emission in the Galactic Center black
hole, Sagittarius~A* (Sgr~A*). Comparing such observations to
time-dependent general relativistic magnetohydrodynamic (GRMHD)
simulations requires statistical tools that explicitly consider the
variability in both the data and the models. We develop here a
Bayesian method to compare time-resolved simulation images to variable
VLBI data, in order to infer model parameters and perform model
comparisons. We use mock EHT data based on GRMHD simulations to
explore the robustness of this Bayesian method and contrast it to
approaches that do not consider the effects of variability. We find
that time-independent models lead to offset values of the inferred
parameters with artificially reduced uncertainties. Moreover,
  neglecting the variability in the data and the models often leads
  to erroneous model selections. We finally apply our method to the early
EHT data on Sgr~A*.
\end{abstract}

\section{Introduction}

The Event Horizon Telescope\footnote{\indent http://www.eventhorizontelescope.org/} (EHT) is a global array of telescopes that performs very-long-baseline interferometry (VLBI) at 1.3 and 0.8 mm wavelengths with unprecedented angular resolution. When the full array of stations from Arizona to the South Pole and from Hawaii to France are incorporated, the longest baselines (14 G$\lambda$ at 0.8 mm) will provide angular resolution better than 20 $\mu$as. The main targets of the EHT are the Galactic Center black hole Sagittarius A* (hereafter Sgr~A*) and the nuclear black hole in the galaxy M87, at the center of the Virgo cluster of galaxies. For both of these sources, the apparent size of the event horizon is larger than the EHT resolution, millimeter emission is expected to originate very close to the horizon, and synchrotron opacity is not expected to obscure the innermost regions, thus allowing a clear and high-resolution view of their event horizons.

The images of the accretion flows around these black holes are expected to exhibit a shadow surrounded by a ring of emission \citep{1973blho.conf..215B, 1979A&A....75..228L, 2000ApJ...528L..13F}, with a size and shape that depends primarily on the black hole mass and only very weakly on its spin \citep{2010ApJ...718..446J}. The shadows of the nuclear black holes of the Milky Way and M87 have apparent diameters of 50 $\mu$as and 36 $\mu$as, respectively, assuming Kerr black holes \citep{2012ApJ...758...30J}. Observations of Sgr~A* with three stations in Hawaii, California, and Arizona measured a source size that is comparable to the event horizon scale \citep{2008Natur.455...78D} and provided evidence for variability of its emission \citep{2011ApJ...727L..36F}. The Sgr~A* observations of \cite{2011ApJ...727L..36F} reported not only the increase of the total flux but also the brightening of the correlated flux density in long baselines. Three-station VLBI observations of M87 detected structure that was identified with the base of the relativistic jet and had a size comparable to that of the black hole shadow \citep{2012Sci...338..355D}. For the rest of this paper, we will focus on Sgr~A*, although our results are very general and can be applied to interferometric observations of any target.

Sgr~A* is the closest supermassive black hole to the Earth. It has long been observed at wavelengths ranging from radio to $\gamma$-rays to study accretion physics around the black hole (see \citealt{2013CQGra..30x4003F} for a recent review). In the near-infrared, adaptive optics observations of orbiting stars provide the mass of and distance to the black hole \citep{2008ApJ...689.1044G, 2009ApJ...707L.114G, 2009ApJ...692.1075G, 2015MNRAS.447..948C}. The angular size of Sgr~A*, set by the ratio of mass to distance, is more accurately measured and is the largest of any known black hole.

Observations at millimeter wavelengths provide the opportunity to understand the accretion process near the event horizon of Sgr~A*. At these wavelengths, the strong interstellar scattering that greatly blurs the image at longer wavelengths \citep[e.g.,][]{2001AJ....121.2610D, 2006ApJ...648L.127B} is reduced and the spectrum of Sgr~A* implies a transition from optically thick to thin emission, providing a clear view of the event horizon \citep[e.g.,][]{1998ApJ...492..554N, 2000ApJ...541..234O, 2006JPhCS..54..354M, 2015ApJ...802...69B}. Sgr~A* is observed to vary on intraday time scales at nearly all wavelengths, including millimeter wavelengths \citep[e.g.,][]{2008ApJ...682..373M, 2009ApJ...706..348Y, 2014MNRAS.442.2797D}, which suggests a highly dynamic environment around the black hole.

There have been many theoretical efforts to simulate accretion onto Sgr~A* \citep[see][for a recent review]{2014ARA&A..52..529Y}. These calculations have generally focused on reproducing its spectrum and measured emission size through both semi-analytic stationary models and time-dependent magnetohydrodynamic (MHD) and general relativistic MHD (GRMHD) simulations. Time-independent models, including \citet{2000ApJ...541..234O}, \citet{2003ApJ...598..301Y}, and \citet{2007MNRAS.379..833H}, have used variations in the magnetic field, electron density, and temperature profiles computed from hydrodynamic models to match observational constraints. Incorporating GR ray-tracing, these models can also be used to constrain properties such as the black hole spin and orientation  \citep[e.g.,][]{2009ApJ...697...45B, 2011ApJ...735..110B}. Time-resolved simulations of the accretion flow can similarly be adjusted through a variety of parameters, prescriptions, and initial conditions to better match existing observations. Such simulations have the potential to capture additional observational details that are not encoded in the stationary simulations, including polarization properties \citep[e.g.,][]{2005ApJ...621..785G, 2012ApJ...755..133S, Gold:2016ud}, and especially the variability of the total emission and source structure. Several (GR)MHD simulations that can match numerous properties of Sgr~A* \citep[e.g.,][]{2009ApJ...701..521C, 2009ApJ...706..497M, 2010ApJ...717.1092D, 2012JPhCS.372a2023D, 2012ApJ...755..133S, 2013A&A...559L...3M, 2015ApJ...812..103C, 2015ApJ...799....1C} now exist in the literature.

The prospect of millimeter-wavelength EHT observations that can resolve Sgr~A* in both space and time, alongside simulations and existing observations that show temporal changes, indicates a need to compare data and models in a time-dependent manner. This should help us constrain the GRMHD model and physical parameters to describe the accretion flow around Sgr~A*. However, there are so far no tools with which time-dependent data can be analyzed in the context of time-variable models. For example, given the likely changes in the structure of the emission region and the stochastic nature of MHD turbulence, a snapshot by snapshot comparison is neither feasible nor meaningful. In this paper, we develop a statistical method to compare interferometric measurements with simulated mock data from time-dependent GRMHD models. In particular, we employ Bayesian tools to compute the posterior likelihood of model parameters given the observational data. We apply this to several simulations and verify the method using mock data. Then, we use EHT observational data from 2007 and 2009 to demonstrate the application of the method on Sgr~A*.

\section{Bayesian data analysis}

In this section, we develop a general Bayesian framework that allows us to compare an ensemble of data points, such as VLBI visibilities, to an ensemble of simulated data predicted by a theoretical model, when both the data and the model exhibit intrinsic variability.

\subsection{Formalism}

We follow the standard Bayesian approach to calculate the posterior likelihood $P(\mathbf{w}|{\rm data}$) that a theoretical model described by a vector of parameters $\mathbf{w}$ is in agreement with a set of data, i.e.,
\begin{equation}
  P (\mathbf{w} \vert \textrm{data}) = C P_{\rm pri}(\mathbf{w}) P ( \textrm{data} \vert \mathbf{w}).
  \label{eq:bayesgen}
\end{equation}
Here, $P_{\rm pri}(\mathbf{w})$ is the prior likelihood over the model parameters and we calculate the proportionality constant $C$ such that the integral of the posterior likelihood over the parameter space is equal to unity.

The vector of parameters for the GRMHD models that will be compared to
EHT observations includes the properties of the black hole spacetime
(its mass, spin, and any deviations from the Kerr solution), geometric
properties of the orientation of the observer (the orientation and
inclination of the spin axis on the observer's sky), and parameters
related to the microphysical properties of the plasma (e.g., the ratio
of electron-to-ion temperatures in phenomenological models). Moreover,
the results of the GRMHD simulations will need to be convolved with a
model that describes the image blurring caused by interstellar
scattering, which itself will have a number of parameters. Even though
we explore below the characteristics of this Bayesian framework using
a simplified two-parameter comparison between simulations and data,
our methods are general and the full parameter space will need to be
considered and explored when fitting actual EHT data.

Hereafter, we will consider a small set for the black hole spin, which has little effect on the overall properties of the black hole image~\citep[see][]{2010ApJ...718..446J,2011ApJ...735..110B}, and fix its inclination with respect to the observer, which can be constrained significantly by the overall size of the 1.3 mm image of the black hole \citep{2015ApJ...798...15P}. We will also consider discrete models for the plasma thermodynamics~\citep[see][]{2015ApJ...799....1C}, with parameters chosen such that the simulations reproduce the broadband spectral properties of Sgr~A*. This set of choices leaves us with two parameters that need to be obtained by comparing models to interferometric data, i.e., an overall multiplication factor for the flux density, $F_{0}$, and the angle $\xi$ between the east-west axis and the projection of the black hole spin angular momentum onto the sky plane, measured in degrees east of north~\citep[e.g.,][]{2011ApJ...735..110B}\footnote{Another often used angle to measure the same orientation is denoted by $\phi$ and is the complementary angle to $\xi$, measured in degrees north of east \citep[e.g.,][]{2009ApJ...697.1741B}; clearly, $\xi = 90^{\circ} - \phi$.}. Under these assumptions, Bayes' theorem becomes
\begin{equation}
  P (\xi, F_{0} \vert \textrm{data}) = C P_{\rm pri} (\xi) P_{\rm pri} (F_{0}) P ( \textrm{data} \vert \xi, F_{0}).
  \label{eq:bayeseq}
\end{equation}
We assume flat prior likelihoods over the two model parameters $P_{\rm pri} (\xi)$ and $P_{\rm pri} (F_{0})$, with $-180^{\circ} \le \xi < 180^{\circ}$ and $F_{\rm min}\le F_0\le F_{\rm max}$, with the range of the overall flux normalization to be specified later.

The last term in Equation~(\ref{eq:bayeseq}), $P ( \textrm{data} \vert \xi, F_{0})$, measures the likelihood that a particular set of data is consistent with a set of model parameters. For the case of EHT, data and simulations need to be compared directly as interferometric observables. A fundamental interferometric observable is the {\it{visibility}}, which is a sample of the Fourier transform of the sky image, $I (\alpha,\beta)$, at discrete spatial frequencies. The visibility, $\mathcal{V}(u,v)$, is defined as
\begin{equation}
  \mathcal{V}(u,v) = \iint I(\alpha,\beta) \exp \left[ - 2 \pi i (u\alpha + v\beta) \right] d\alpha d\beta
\end{equation}
for a projected antenna separation of ($u,v$) wavelengths within the plane perpendicular to the line of sight.

At any given observing epoch, the EHT will generate a set of simultaneous visibility amplitudes at many baselines as well as sets of closure phases on baseline triangles and closure amplitudes. For the purposes of this initial investigation, and to simplify our notation, we will assume that we only have observations of visibility amplitudes and that each measurement has independent uncertainties from the others. In other words, we will assume that
\begin{equation}
  P(\textrm{data} \vert \xi, F_{0}) \equiv \prod\limits_{i=1}^M \prod\limits_{j=1}^{N} P_{ij}({\rm data}\vert \xi, F_{0}),
  \label{eq:likeprod}
\end{equation}
where $i=1, 2, \ldots, M$ and $j=1, 2, \ldots, N$ denote baselines and epochs, respectively.

\begin{figure}[t]
\begin{center}
\includegraphics[width=8.5cm]{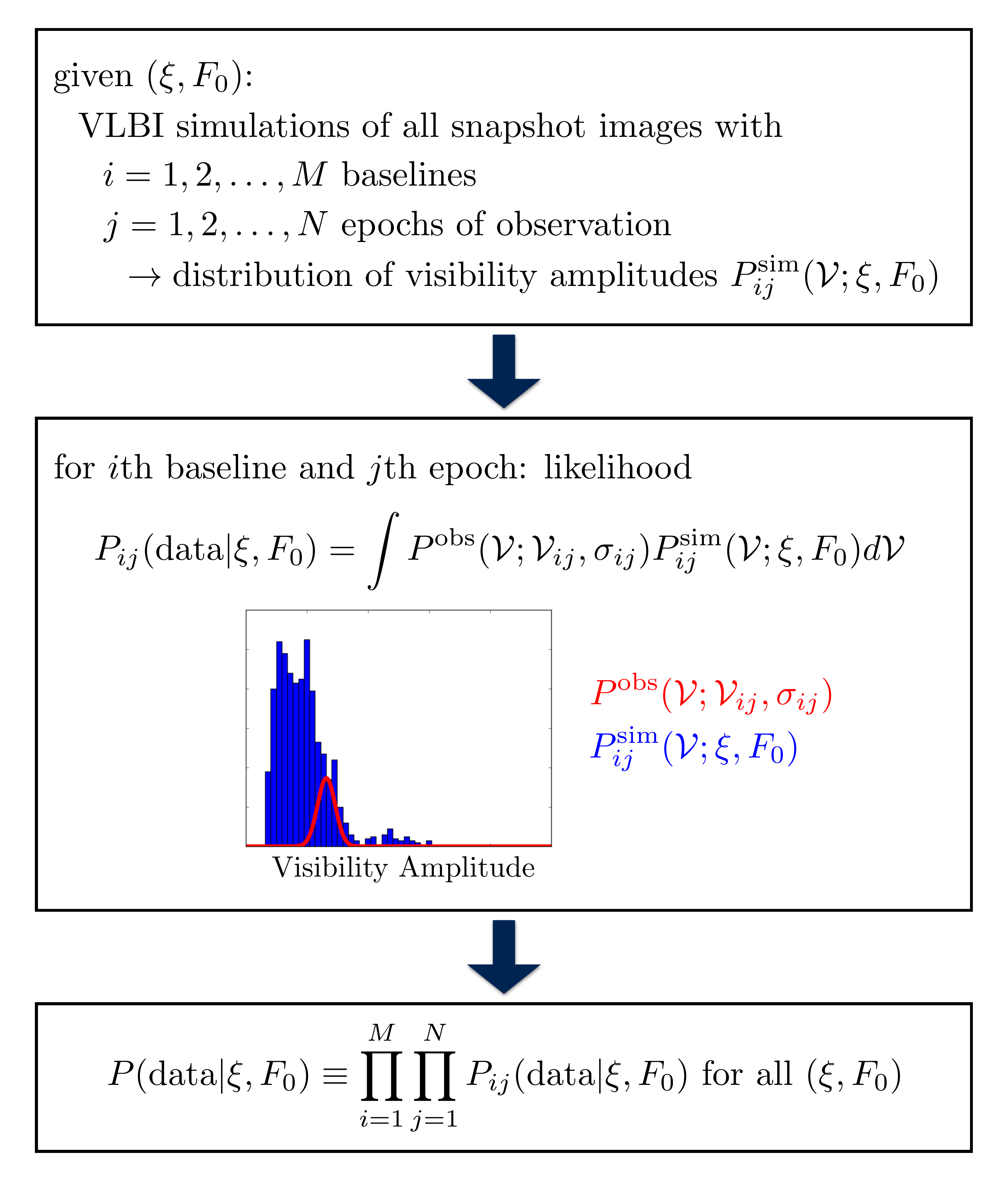} 
\end{center}
\caption{Flow chart of the algorithm to test the performance of the Bayesian data analysis.}
\label{fig:flowchart}
\end{figure}

Each observation is characterized by a likelihood $P^{\rm obs}({\cal V};{\cal V}_{ij},\sigma_{ij})$, centered at ${\cal V}_{ij}$ and with a dispersion of $\sigma_{ij}$. If, for any pair of model parameters $(\xi,F_0)$, the model made a single prediction for the visibility amplitude for baseline $i$ and epoch $j$, say ${\cal V}^{\rm sim}_{ij}(\xi,F_0)$, then
the single-baseline, single-epoch likelihood in Equation~(\ref{eq:likeprod}) would simply be equal to
\begin{equation}
  P_{ij}({\rm data}\vert \xi, F_{0})=P^{\rm obs}[{\cal V}^{\rm sim}_{ij}(\xi,F_0);{\cal V}_{ij},\sigma_{ij}]\;.
\end{equation}
However, GRMHD models are highly variable~\citep[e.g.,][]{2015ApJ...812..103C} and, for each baseline, they predict a range of visibility amplitudes, with a likelihood that we denote by $P^{\rm sim}_{ij} (\mathcal{V}; \xi, F_{0})$. In this case, we write the posterior likelihood that a particular observation is consistent with the model predictions as
\begin{equation}
  P_{ij} ({\rm data}\vert \xi, F_{0}) = \int P^{\rm obs}(\mathcal{V}; {\mathcal{V}}_{ij}, {\sigma}_{ij}) P^{\rm sim}_{ij} (\mathcal{V}; \xi, F_{0}) d {\mathcal{V}}\;.
  \label{eq:likelihoodeq}
\end{equation}
Our goal is to find the set of model parameters ($\xi$ and $F_0$ in the example here) that maximizes the posterior likelihood in Equation~(\ref{eq:likeprod}), given a set of observations. This Bayesian analysis process, presented in Figure~\ref{fig:flowchart} as a flow chart, selects models for which the distribution of predicted visibilities in each baseline matches the distribution of observed visibilites.

The first factor in Equation~(\ref{eq:likelihoodeq}), $P^{\rm obs}(\mathcal{V}; {\mathcal{V}}_{ij}, {\sigma}_{ij})$, is well
understood. The real and imaginary components of the complex visibility individually follow Gaussian distributions. The visibility amplitude is the magnitude of the complex vector described by those variables, and its posterior likelihood is not Gaussian. The appropriate posterior likelihood for the visibility amplitude, known as a Rice distribution, is
\begin{equation}
  P^{\rm obs} (\mathcal{V}; {\mathcal{V}}_{ij}, {\sigma}_{ij}) = \frac{\mathcal{V}}{\sigma_{ij}^2} \exp \left[ - \frac{ \left(\mathcal{V}^2 + {\mathcal{V}}_{ij}^2 \right)}{2 {\sigma}_{ij}^2} \right] {I_0} \left( \frac{\mathcal{V} {\mathcal{V}}_{ij}}{\sigma_{ij}^2} \right),
\end{equation}
where $I_0$ is the zeroth order modified Bessel function of the first kind. This likelihood approaches a Gaussian for large signal-to-noise ratios \citep[see Chapter 6 of][]{TMS2001}.

In order to simplify the calculation of integral~(\ref{eq:likelihoodeq}), we make use of the fact that the distribution of simulated visibility amplitudes $P^{\rm sim}_{ij} (\mathcal{V} ; \xi, F_{0})$ has discrete values rather than a continuous distribution, as it is obtained from snapshot images. For $T$ snapshots, 
\begin{equation}
  P^{\rm sim}_{ij} (\mathcal{V} ; \xi, F_{0}) = \frac{1}{T} \sum\limits_{t=1}^{T} \delta \left[\mathcal{V} - \mathcal{V}^{\rm sim}_{ij}(t; \xi, F_{0}) \right]
\end{equation}
where $\mathcal{V}^{\rm sim}_{ij}(t; \xi, F_{0})$ is the visibility amplitude sampled at $(u_{ij}, v_{ij})$ coordinates from the $t$th snapshot with the given set of parameters $(\xi, F_{0})$. Therefore, Equation~(\ref{eq:likelihoodeq}) simplifies to
\begin{eqnarray}
&&P_{ij} ({\rm data} \vert \xi, F_{0}) = \nonumber\\
&&\quad\frac{1}{T} \sum\limits_{t=1}^{T} \frac{\mathcal{V}^{\rm sim}_{ij}(t; \xi, F_{0})}{{{\sigma}_{ij}}^2} \exp \left[ - \frac{ (\mathcal{V}^{\rm sim}_{ij}(t; \xi, F_{0}) + {\mathcal{V}}_{ij})^2}{2 {{\sigma}_{ij}}^2} \right] \nonumber\\
&& \qquad\times {I_0} \left( \frac{\mathcal{V}^{\rm sim}_{ij}(t; \xi, F_{0}) {\mathcal{V}}_{ij}}{{\sigma_{ij}}^2} \right)
\label{eq:discreteprob}
\end{eqnarray}
for the likelihood $P^{\rm sim}_{ij} (\mathcal{V} ; \xi, F_{0})$, with given parameters.

In the following set of examples, we calculate the posterior likelihoods over the model parameters using Equations~(\ref{eq:bayeseq}), (\ref{eq:likeprod}), and (\ref{eq:discreteprob}). In real applications, however, observations
on different baselines are performed simultaneously, so we must consider a joint probability distribution in visibility space. For example, when an observation is carried out with three stations (or equivalently, three baselines), the posterior likelihood for the observations, $P^{\rm obs} (\mathcal{V})$, becomes a multivariate Rician. Moreover, we need to consider the posterior likelihood, $P^{\rm sim} (\mathcal{V})$, that the model predicts a particular combination
of simultaneous visibilities on the same baselines.

Finally, we can trivially incorporate to our Bayesian inference other interferometric observables, such as closure phases and closure amplitudes. The closure phase is the sum of visibility phases around the triangle of baselines formed by any set of three stations. The closure phase is independent of instrumental and atmospheric phase fluctuations and depends only on the visibility phase of the source. If we have $\mathcal{N}$ stations where $\mathcal{N}\geq 3$, ${_{\mathcal{N}-1}}C_2 = (\mathcal{N} - 1) (\mathcal{N} - 2) / 2$ independent closure phases can be measured. The posterior likelihood for closure phase data becomes
\begin{equation}
  P_{lj}({\rm data} \vert \xi, F_{0}) = \int P^{\rm obs}(\Phi; {\Phi}_{lj}, {\sigma}_{lj}) P^{\rm sim}_{lj} (\Phi; \xi, F_{0}) d\Phi
\end{equation}
for $l=1, 2, \ldots, {_{\mathcal{N}-1}}C_2$ sets of telescopes (closure phase ``triangles''), and $j=1, 2, \ldots, N$ observations. Here, $P^{\rm obs}(\Phi; {\Phi}_{lj}, {\sigma}_{lj})$ and $P^{\rm sim}_{lj} (\Phi; \xi, F_{0})$ are the likelihoods for closure phases from the observation and simulations, similar to $P^{\rm obs} (\mathcal{V}; {\mathcal{V}}_{ij}, {\sigma}_{ij})$ and $P^{\rm sim}_{ij}(\mathcal{V} ; \xi, F_{0})$ in Equation~(\ref{eq:likelihoodeq}).

\subsection{Model Comparison}\label{modelcomp}

In addition to inferring the most likely parameters of a given model from observations, we are also interested in comparing the ability of different models to describe the observed distributions of visibility amplitudes and closure quantities.

Information criteria approaches such as the Akaike Information Criterion (AIC) and the Bayesian Information Criterion (BIC) have been adopted elsewhere to evaluate the quality of astrophysical models \citep[e.g.,][]{2007MNRAS.377L..74L}. \cite{2011ApJ...735..110B} applied these criteria in evaluating the relative success of time-independent models of Sgr~A* in fitting EHT data. The BIC is defined as
\begin{equation}
  \textrm{BIC} \equiv -2 \log {\mathcal{L}}_{\textrm{max}} + k  \log N,
  \label{eq:biceq}
\end{equation}
where ${\mathcal{L}}_{\textrm{max}}$ is the maximum likelihood using the parameters within the model, $k$ is the number of free parameters, and $N$ is the number of data points. The definition implies that the smaller BIC value suggests a better fit to the data. However, criteria such as the BIC consider only the maximum posterior likelihoods and
do not account for the relative extent of the volumes in parameter space within which each model is consistent with the data.

In the context of our work, we will perform model comparisons in terms of an appropriately defined Bayesian evidence. If we consider each model as a single point in a discrete parameter space, then, given the data, we can use Bayes' theorem to calculate the posterior likelihood for each model as
\begin{equation}
  P(\mathcal{M}_{m} \vert \textrm{data}) = C P_{\rm pri}
  (\mathcal{M}_{m}) P(\textrm{data} \vert \mathcal{M}_{m}),
  \label{eq:models_gen}
\end{equation}
Here $m$ is the discrete index specifying each model \{$\mathcal{M}_{m}$\}, which is governed by a set of parameters
$\mathbf{w}_m$.  If we denote by $P_{\rm pri}(m)$ the prior likelihood for each model and by $P_{{\rm pri},m}({\mathbf w}_m)$ the prior likelihoods for each set of parameters within model $m$, then we can write
\begin{equation}
  P(\mathcal{M}_{m} \vert \textrm{data}) = C P_{\rm pri}(m)
  P_{{\rm pri},m}(\mathbf{w}_{m}) P_m(\textrm{data} \vert \mathbf{w}_{m})\;.
  \label{eq:models_spec}
\end{equation}
The last term in Equation~(\ref{eq:models_spec}) is the posterior Bayesian likelihood of the data, given a model $m$ and a set of model parameters, e.g., Equation~(\ref{eq:likeprod}).

In order to calculate the Bayesian evidence for model $m$, we marginalize Equation~(\ref{eq:models_spec}) over the space of model parameters, i.e.,
\begin{equation}
  {\cal L}(m \vert \textrm{data}) = C P_{\rm pri}(m)\int P_{{\rm pri},m}(\mathbf{w}_{m}) P_m(\textrm{data} \vert \mathbf{w}_{m}) d\mathbf{w}_m.
\end{equation}
Assuming a flat prior of models and the same prior likelihood over all model parameters for each model, this intergral simplifies to
\begin{equation}
 \mathcal{L} (m\vert\textrm{data}) = C^\prime\int P_m(\textrm{data} \vert \mathbf{w}_m) d\mathbf{w}_m,
 \label{eq:evidence}
\end{equation}
where $C^\prime$ is an appropriate normalization constant.

The integrated likelihood $\mathcal{L} (m\vert \textrm{data})$ provides a measure of model selection statistics.  As is the standard practice in Bayesian inference, we use the Bayesian evidence only for comparison among models and not to assign a posterior likelihood to any model individually.

\section{Likelihood with mock data}

We now turn our attention to the application of the statistical method described in the previous section to mock and real EHT data. To this end, we make use of a set of previously studied GRMHD simulations labeled A $\rightarrow$ E, for which time-dependent spectra and images were computed and fit to the time-averaged properties of Sgr~A* \citep{2015ApJ...799....1C}.

\subsection{GRMHD Simulations}

In a recent study, \cite{2015ApJ...812..103C} investigated the variability and millimeter/infrared flare characteristics of GRMHD simulations. The simulation of the accretion flow was performed with the 3-dimensional GRMHD code \texttt{HARM} \citep{2003ApJ...589..444G, 2006MNRAS.368.1561M, 2009MNRAS.394L.126M, 2012MNRAS.426.3241N, 2013MNRAS.436.3856S} and the ray-tracing of photon trajectories was performed with the fast GPU-based algorithm \texttt{GRay} \citep{2013ApJ...777...13C}. The simulated Sgr A* images with horizon-scale resolution and their spectra were fit to both the observed broadband spectra and the overall 1.3 mm source size determined by EHT observations \citep{2015ApJ...799....1C}.

\cite{2015ApJ...812..103C, 2015ApJ...799....1C} simulated high cadence images of the accretion flow around Sgr~A* for each model. The simulation results were recorded with a timestep of $10 GM{c}^{-3}$, which corresponds to 212 s for the mass of Sgr~A*. Each snapshot contains four levels of field of view: $16 M \times 16 M$, $64 M \times 64 M$, $256 M \times 256 M$, and $1024 M \times 1024 M$, containing $512 \times 512$ pixels. Here $1 GMc^{-2}$ corresponds to 5.1 $\mu$as for the distance to Sgr~A*. We use a composite of the multi-level resolutions for our analysis in order to encompass all the emission near and away from the black hole and to generate interferometric visibilities. It is reasonable to limit our images to $1024 M \times 1024 M$, because the corresponding 10.2 $\mu$as pixel scale is considerably finer than the highest spatial frequency (58.9 $\mu$as fringe spacing) sampled by the EHT data we use as a comparison.

The GRMHD models of \citet[see also \citealp{2012MNRAS.426.3241N, 2013MNRAS.436.3856S}]{2015ApJ...812..103C, 2015ApJ...799....1C} were categorized into two classes based on initial magnetic field configurations: Standard And Normal Evolution (SANE, disk-dominated) and Magnetically Arrested Disk (MAD, jet-dominated) models, which use multi-loop and single-loop initial magnetic fields, respectively. The simulation images at 1.3 mm for these two classes of models show dissimilar features from each other: in SANE models, the emission originates from a crescent-like shape in the disk region, whereas it is concentrated at the footprints of the outflowing material in the MAD models. \cite{2015ApJ...812..103C, 2015ApJ...799....1C} also explored the dependence of the predictions of the GRMHD models on the black hole spin parameter $a$ and on the way in which the electron temperature was prescribed.

\begin{figure}[t]
\begin{center}
\includegraphics[width=7cm]{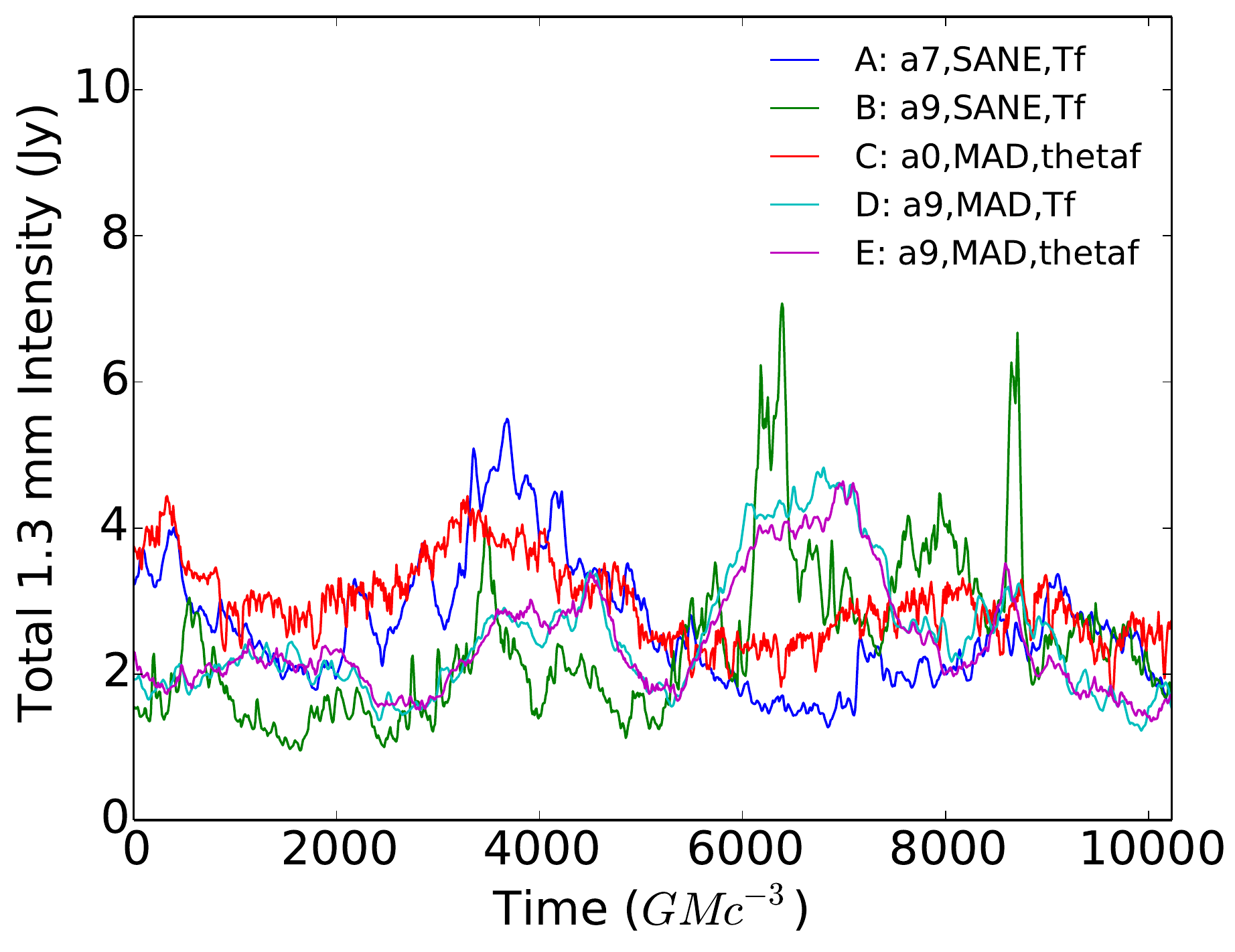} 
\end{center}
\caption{Light curves of Models A to E at 1.3 mm. The total fluxes of each snapshot are plotted with a 10 $GM c^{-3}$ timestep, over a 10,000 $GM c^{-3}$ duration. The combination $GM c^{-3}$ corresponds to 21.2 s for the mass of Sgr~A*. SANE models not only show larger variability than MAD models, but also often display sudden increases of the total intensity, i.e., flares.}
\label{fig:lightcurve}
\end{figure}
In some of their models, the electron temperature in the funnel region of the accretion flow was fixed to be constant (constant $T_{\textrm{\tiny{e,funnel}}}$). In other models, the ratio of the electron and ion temperatures in the funnel was fixed (constant ${\theta}_{\textrm{\tiny{funnel}}}$). In both cases, the electron and ion temperatures have a constant ratio in the disk. Among all the configurations they explored, they narrowed the possibilities down to five models that account for both the broadband spectrum of Sgr~A* and its overall image size at 1.3 mm:
\begin{itemize}
\setlength\itemsep{0em}
\item Model A: a=0.7, SANE, constant $T_{\textrm{\tiny{e,funnel}}}$
\item Model B: a=0.9, SANE, constant $T_{\textrm{\tiny{e,funnel}}}$
\item Model C: a=0.0, MAD, constant ${\theta}_{\textrm{\tiny{funnel}}}$
\item Model D: a=0.9, MAD, constant $T_{\textrm{\tiny{e,funnel}}}$
\item Model E: a=0.9, MAD, constant ${\theta}_{\textrm{\tiny{funnel}}}$
\end{itemize}
All models show variability over the entire length of the simulation, i.e., over $\sim$ 10,000 $GM{c}^{-3}$ ($\sim$ 60 hrs). Figure~\ref{fig:lightcurve} shows the 1.3 mm light curves for the simulations. Overall, the SANE models are observed to produce more variability than the MAD models.

We generated visibilities from the simulations using our own VLBI simulation code. The algorithm calculates $(u,v)$ coordinates from the observing time and celestial coordinates of the source, performs Fourier transform of the input image, and samples complex visibilities according to the telescope array setup. We tested the results of this code against the MIT Array Performance Simulator\footnote{http://www.haystack.mit.edu/ast/arrays/maps/index.html} (MAPS) and found excellent agreement between the simulated visibilities.

\begin{figure*}
\begin{center}
\includegraphics[width=16.5cm]{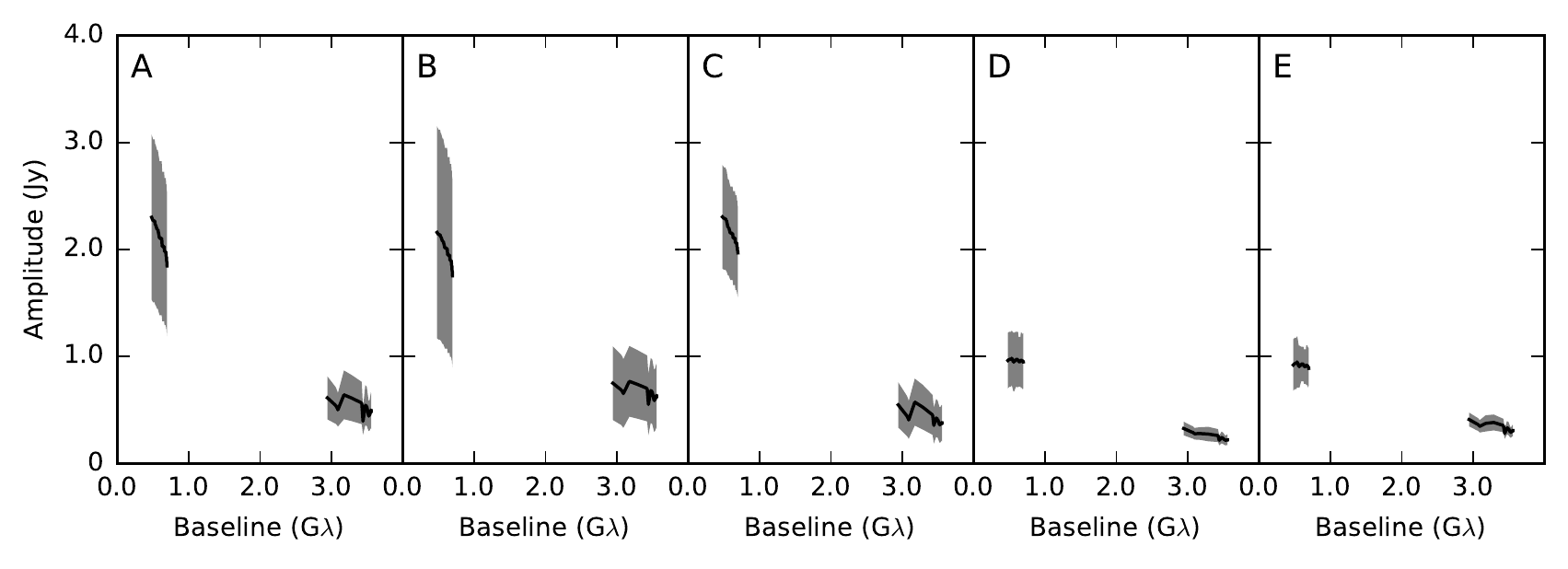} 
\end{center}
\caption{Sample VLBI visibility amplitudes of Models A to E. We generated the simulated visibility amplitudes using three stations in Hawaii, California, and Arizona. We produced the visibility amplitudes from GRMHD images, using $(u,v)$ coordinates of the EHT observation data in 2007 and 2009, without rotation ($\xi=0^{\circ}$) and flux scaling ($F_0 = 1$). For all $(u,v)$ coordinate pairs, we plot the mean visibility amplitudes of 1024 snapshots in solid line as well as their $\pm1 \sigma$ range in the shaded region.}
\label{fig:vis_variability}
\end{figure*}

Figure~\ref{fig:vis_variability} shows the range of predicted visibility amplitudes for the five GRMHD models. The SANE models show more variability in the visibility space than the MAD models, just as they do in total intensity (Figure~\ref{fig:lightcurve}). This variability is explored in more detail in the companion paper, \cite{2016arXiv160106799M}.

In the following sections, we describe tests of our Bayesian analysis using mock data from these simulations. These tests demonstrate that our statistical method successfully recovers model parameters (section~\ref{paramselect}) and can faithfully distinguish between models (section~\ref{modelselect}).

\subsection{Parameter estimation}\label{paramselect}

In our first test we investigated the performance of our statistical method in inferring model parameters. We constructed the test as follows:
\begin{itemize}
\setlength\itemsep{0em}
\item[--] We assumed a fiducial set of model parameters ($\xi$, $F_0$) = ($-30^{\circ}$, $1$).
\item[--] We generated data using three baselines with three stations: Hawaii - Arizona, Hawaii - California, and Arizona - California.
\item[--] We sampled $(u,v)$ coordinates during the period when three stations can perform simultaneous observations of Sgr~A*.
\item[--] We generated visibilities using model D, which is a MAD model.
\item[--] We randomly selected visibility amplitudes from the distribution of simulated visibilities, $P_{ij}^{\textrm{sim}}(\mathcal{V}; -30^{\circ}, 1)$, for each baseline. Visibilities were taken from the same simulation frame for all three baselines. We then added random Gaussian noise to the complex visibilities to produce mock observational data from the visibility amplitude samples.
\item[--] We computed the posterior likelihood $P( \textrm{data} \vert
  \xi, F_{0})$ using simultaneous observations on three baselines, as
  described in Equation~(\ref{eq:likeprod}). We considered
    orientation $\xi$ in the range from -90$^{\circ}$ to 90$^{\circ}$,
    as visibility amplitudes without phase information are degenerate
    under rotations of 180$^{\circ}$. We repeated the calculation for
  5000 different realizations of the mock data.
\end{itemize}

\begin{figure*}
\begin{center}
\includegraphics[width=19cm]{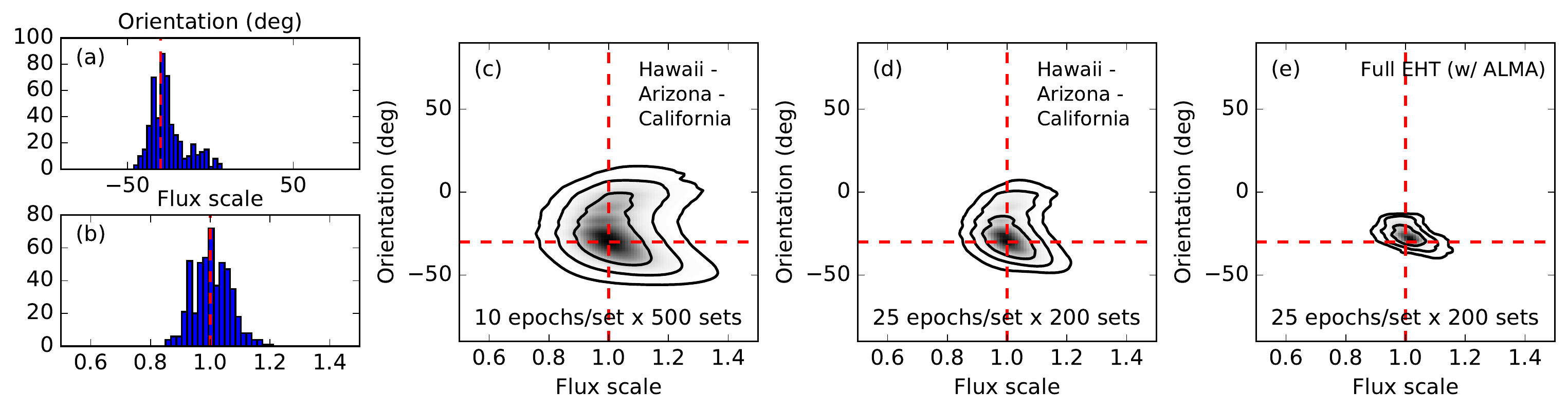} 
\end{center}
\caption{Parameter estimation test with mock data generated using model D and assuming a black hole spin orientation of $\xi = -30^{\circ}$ and an overall flux normalization of $F_0 = 1$. The first two panels show the one-dimensional likelihoods for 500 realizations of observations, each containing 10 epochs over (a) the orientation ($\xi$) and (b) flux scale ($F_0$), evaluated at cross sections of the two-dimensional parameter space that encompasses the point of maximum likelihood. Panels (c) and (d) show the averaged likelihoods in parameter space for (c) 500 realizations each containing 10 epochs and (d) 200 realizations each containing 25 epochs. Maximum likelihoods occur at $\xi = -30^{\circ}$ and $F_0 = 1$ for both cases. Contours indicate the 68.3\%, 95.5\%, and 99.7\% credible regions. Given a realistic observational setup, our Bayesian method recovers the assumed parameters with no biases. Panel (e) is similar to (d), but with more complete coverage of a six-station EHT (see section~\ref{section:future_eht}).}
\label{fig:param_mock_D}
\end{figure*}

We averaged the posterior likelihoods to form 500 sets of data, where each set contains $N=10$ observations. For every set of data, we found the orientation and flux scale where the maximum likelihood occurs. The distributions of these maximum likelihood parameters for our input simulation are shown in Figure~\ref{fig:param_mock_D}(a) and (b). The red vertical lines mark the model parameters ($\xi$, $F_0$) that were assumed when generating the mock visibility data. The histograms indicate that the orientation and flux scale of the highest likelihoods of all data sets are distributed around the assumed parameters. We do not see any bias in the inferred parameters.

Figure~\ref{fig:param_mock_D}(c) shows the averaged posterior likelihoods of 500 sets of data. The red dotted lines in the parameter space again mark the combination of the assumed model parameters ($\xi$, $F_0$) = ($-30^{\circ}$, $1$). The lines cross very near the peak of the posterior likelihood, and their intersection is located inside the 68\% credible region of the likelihood. Figure~\ref{fig:param_mock_D}(d) shows the averaged likelihoods when the 5000 mock visibilities are grouped into 200 sets, each including $N=25$  observations. Compiling a larger number of observations causes the allowed region of the parameter space to contract, as expected.

\subsection{Effect of Future EHT Stations}
\label{section:future_eht}
In the near future, more stations will be incorporated into the EHT, such as the Atacama Large Millimeter/submillimeter Array (ALMA) and the South Pole Telescope (SPT). We performed a mock observation test to explore the effect of future EHT stations. The test is similar to that of the previous section, but assuming we have 15 baselines with 6 stations. The stations are the Submillimeter Array (SMA) in Hawaii, the Combined Array for Research in Millimeter-wave Astronomy (CARMA) in California, the Submillimeter Telescope (SMT) in Arizona, the Large Millimeter Telescope (LMT) in Mexico, ALMA in Chile, and the SPT at the South Pole. The sampled $(u,v)$ coordinate sets include non-simultaneous observations because considering only simultaneous detections for six stations greatly limits the observing period. Visibility amplitude errors are estimated from the baseline sensitivity, which is proportional to the geometric mean of the System Equivalent Flux Densities (SEFDs) of antennas.\footnote{SMA phased array: 4000 Jy,  CARMA: 10,000 Jy, SMT: 11,000 Jy, LMT: 1400 Jy, ALMA phased array: 100 Jy, SPT: 9000 Jy.} We assumed 10 s coherent integration time and 2 GHz bandwidth for the sensitivity.

\begin{figure*}
\begin{center}
\includegraphics[width=14cm]{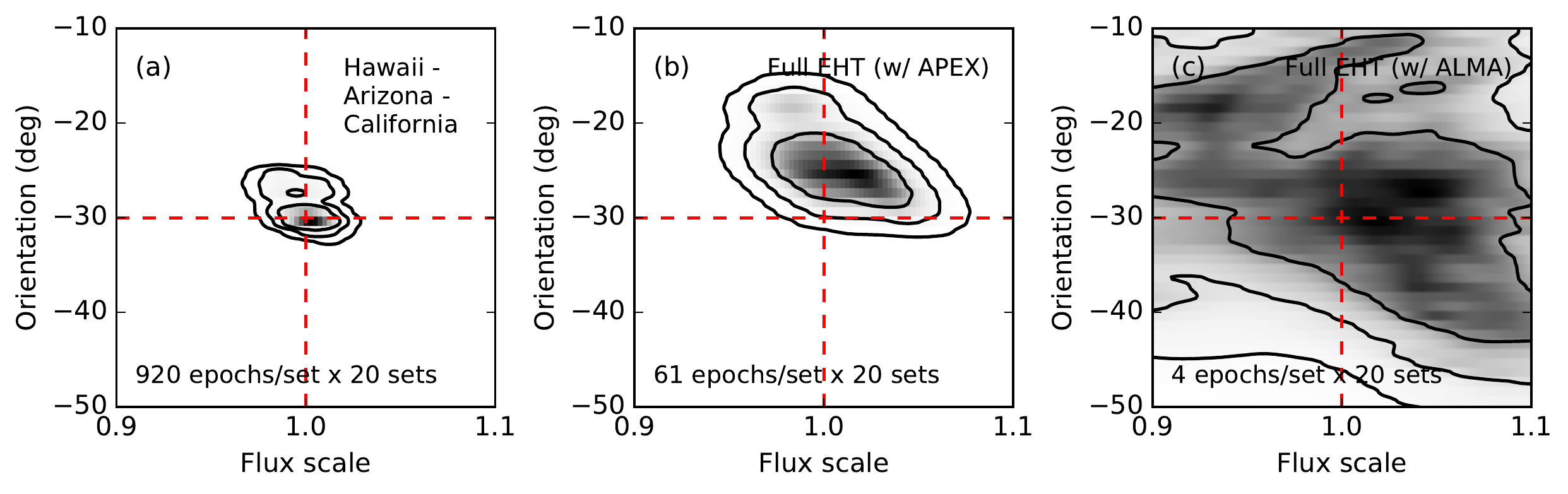} 
\end{center}
\caption{Parameter estimation test with the same underlying parameters as Figure~\ref{fig:param_mock_D}. The three panels show the averaged likelihoods for 20 realizations each in three separate setups. The number of observations in each setup is chosen such that the raw overall sensitivity is held fixed across the three setups. The three panels correspond to (a) 920 epochs on the SMA-CARMA-SMT baseline, where previous EHT observations were done; (b) 61 epochs on baselines including SMA, CARMA, SMT, LMT, APEX, and SPT; (c) 4 epochs including the same baselines as panel (b) but with ALMA replacing APEX. These figures show that sampling the source variability is more important for parameter estimation than the raw sensitivity of the array or the $(u,v)$ coverage.}
\label{fig:futureeht}
\end{figure*}

There are three factors that affect the allowed parameter ranges in the Bayesian analysis: the overall sensitivity of the interferometric array, $(u,v)$ coverage in the visibility space, and sampling of the intrinsic variability of the source. Figure~\ref{fig:param_mock_D}(e) shows the combined effect of these: the increased sensitivity and $(u,v)$ coverage of a six-station EHT will narrow the allowed parameter space compared to Figure~\ref{fig:param_mock_D}(d), which has the same number of observations but fewer stations and lower sensitivity. To assess the importance of source variability in parameter estimation, we constructed three array setups as follows:
\begin{itemize}
\setlength\itemsep{0em}
\item[(a)] The early EHT array of three stations at Hawaii, Arizona, and California, with 920 observing epochs.
\item[(b)] The full EHT array with six stations when the sensitivity of ALMA is replaced with that of the co-located Atacama Pathfinder Experiment (APEX) telescope,\footnote{SEFD of 3600 Jy} with 61 observing epochs.
\item[(c)] The full EHT array with six stations, including ALMA, with four observing epochs.
\end{itemize}
In these three scenarios, we have varied the number of observing epochs such that the accumulated point source sensitivity of the array is constant. That is, the total thermal noise is the same in all the arrays, after accounting for the varying number of observing epochs. The comparison of the six-station array with APEX is in interesting counterpoint to the case with ALMA because it holds the $(u,v)$ coverage fixed (and the sensitivity), but varies the number of observing epochs. Figure~\ref{fig:futureeht} shows that the parameter constraints are much tighter when we average over more observing epochs, emphasizing the importance of sampling the intrinsic source variability when attempting to constrain the model parameters with EHT observations. Figure~\ref{fig:futureeht}(a) has the poorest $(u,v)$ coverage by far, but achieves the tightest constraint, while fixing both $(u,v)$ coverage and sensitivity in Figures \ref{fig:futureeht}(b) and \ref{fig:futureeht}(c) again show that simply sampling more epochs will tighten the parameter constraints significantly.

\subsection{Effect of Time Averaging}
\begin{figure*}
\begin{center}
\includegraphics[width=12cm]{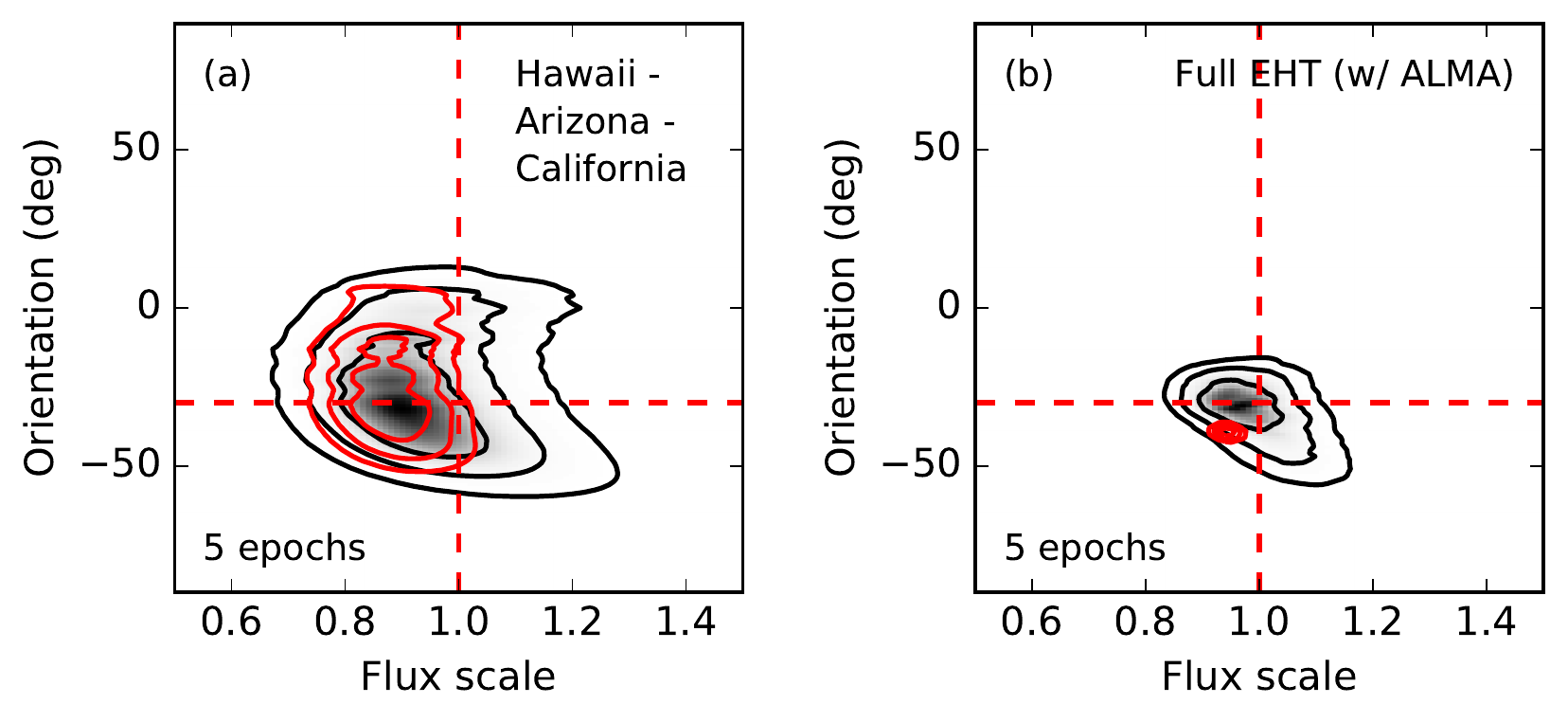} 
\end{center}
\caption{Effect of time averaging in the parameter estimation test for (a) the early EHT baselines and (b) the full EHT baselines including ALMA. We use the identical parameters for the mock data as Figure~\ref{fig:param_mock_D}. The likelihoods of the averaged GRMHD image are overplotted in red contours. The likelihood using the averaged images underestimates the uncertainties of the inferred parameters. The best-fit parameters derived from the time-dependent and the averaged images are offset from each other as well.}
\label{fig:param_mock_D_avg}
\end{figure*}

In the parameter estimation test, we calculated likelihoods
  over parameter space with both time-dependent and averaged GRMHD
  images to compare the effect of time
  averaging. Figure~\ref{fig:param_mock_D_avg} contrasts the
  likelihoods of time-resolved (black contours) and time-independent (red
  contours) inputs to our Bayesian method when the same number of mock
  observations are analyzed. Figures~\ref{fig:param_mock_D_avg}(a) and
  \ref{fig:param_mock_D_avg}(b) are the likelihoods assuming the early
  three-station baselines and the six-station EHT with ALMA,
  respectively. The figure shows that a much smaller region of parameter space is
  allowed when comparing the mock data to a time-averaged image.
  However, the best-fit parameters in the time-averaged case are offset
  from the true parameters, as is most clearly evident in Figure~\ref{fig:param_mock_D_avg}(b).
  This demonstrates the false certainty implied by the small allowed parameter region.
  The time-resolved analysis, by contrast, delivers allowed parameter ranges that are consistent
  with the input simulation.

\subsection{Model selection}\label{modelselect}

We now use the Bayesian method to identify the relative posterior likelihood among several models, as discussed in section~\ref{modelcomp}. For this test we generated mock visibilities following the same procedure as the test of the previous section. We used $(u,v)$ coordinates identical to those in the EHT observations of 2007 and 2009. As discussed earlier, SANE models present greater variability than MAD models (Figure~\ref{fig:vis_variability}). Therefore, we generated two mock data sets, one from model A (SANE) and one from model E (MAD), and performed our model selection test twice (labeled test A and test E, respectively).

We present in Table~\ref{table:modeltest} the marginal likelihoods in Equation~(\ref{eq:evidence}) and the $\triangle$BIC values in Equation~(\ref{eq:biceq}). We obtained these values by fitting the mock model A data and mock model E data with all five models. The ratio of the marginal likelihoods for two models is the Bayes factor, and it measures the relative strength of the evidence against another one. Jeffreys's scale provides a guidance for the interpretation of the Bayes factor \citep{Kass:1995eh}. A ratio between 10 and 100 is translated as strong, and a ratio greater than 100 is regarded as decisive evidence. A similar approach using the difference of information criteria exists and $\triangle$IC greater than 5 and 10 are understood as strong and decisive, respectively.
\begin{deluxetable}{ccccc}
\tablecolumns{5}
\tablewidth{0pc}
\tablecaption{Marginal likelihoods of model estimation tests}
\tablehead{\colhead{} & \multicolumn{2}{c}{Test A} & \multicolumn{2}{c}{Test E}\\
\cline{2-3} \cline{4-5}\\
\colhead{} & \colhead{$\mathcal{L} (m\vert\textrm{data})$} & \colhead{$\triangle$BIC} & \colhead{$\mathcal{L} (m\vert\textrm{data})$} & \colhead{$\triangle$BIC}}
\startdata
Model A & $1.00$ & 0.0 & $5.36 \times 10^{-7}$ & 28.5\\
Model B & $0.799$ & 1.5 & $5.31 \times 10^{-10}$ & 42.8\\
Model C & $7.20 \times 10^{-9}$ & 37.5 & $9.90 \times 10^{-4}$ & 12.7\\
Model D & $1.81 \times 10^{-4}$ & 17.3 & $3.47 \times 10^{-5}$ & 20.8\\
Model E & $2.98 \times 10^{-10}$ & 41.4 & $1.00$ & 0.0
\enddata
\tablecomments{Test A uses mock data using model A, and test E uses mock data using model E. The marginalized likelihoods, $\mathcal{L} (m\vert\textrm{data})$ are normalized to the best model for both tests, as are the BIC values.}
\label{table:modeltest}
\end{deluxetable}

In each test, our analysis prefers the model from which the data were generated. In Test A, model B is only slightly disfavored compared to model A.  This results from the similar shapes of the visibility amplitude distributions in the SANE simulations. The fact that model B has a different spin than model A is also an indication that black hole spin has a very small effect on the overall appearance of SANE models for Sgr A*. In this test, the MAD models are decisively rejected, as expected, because the smaller variability of the model visibilities in the MAD simulations is inconsistent with the mock data. Conversly, Test E decisively favors model E but rejects all four other models.


\section{Applications to early EHT data}

\begin{figure*}[p]
\begin{center}
\includegraphics[width=13.5cm]{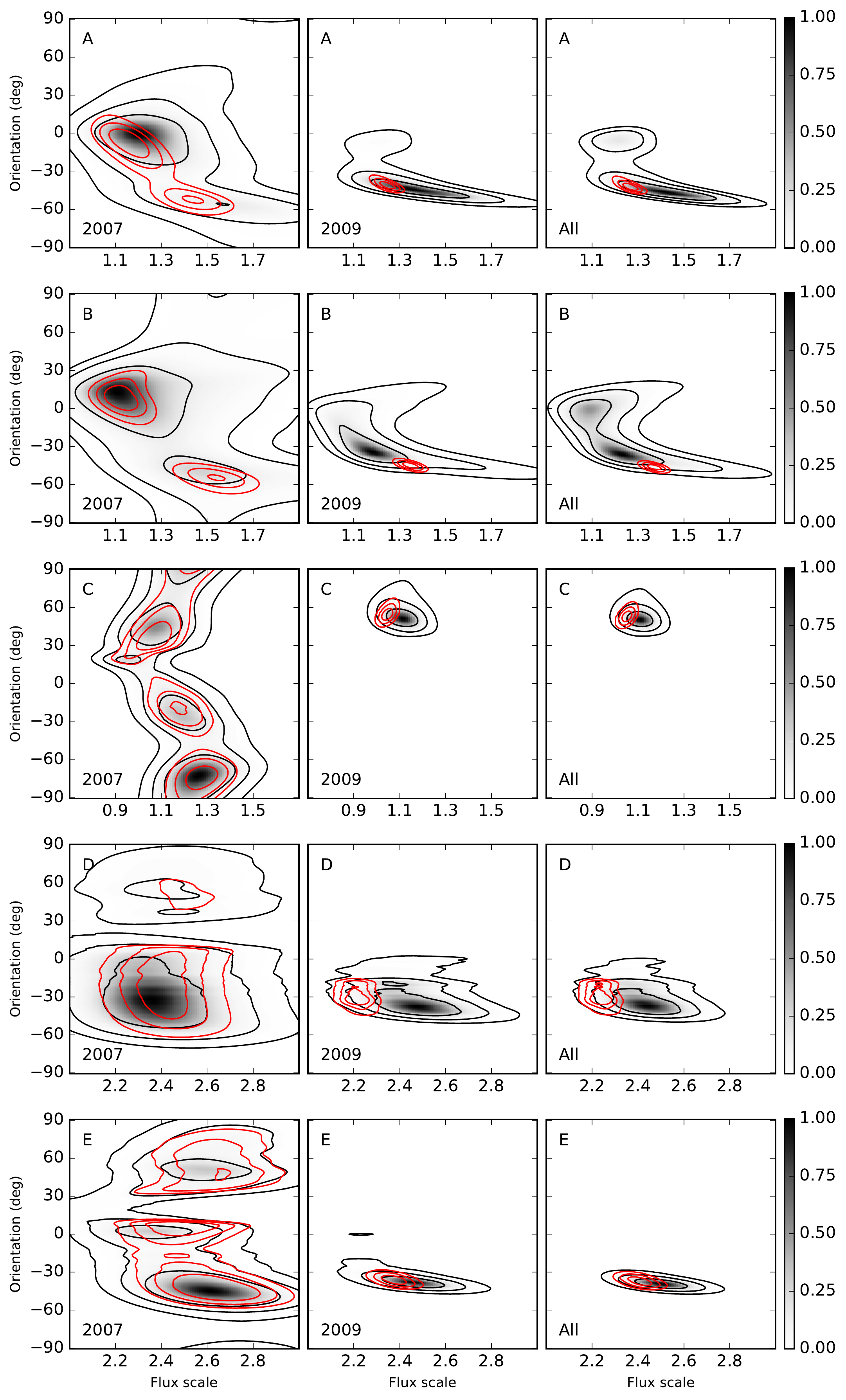}
\end{center}
\caption{Normalized posterior likelihoods over the black hole spin orientation
  $\xi$ and the overall flux normalization $F_0$, when the five GRMHD
  models A to E of \cite{2015ApJ...812..103C} are compared to two
  epochs of EHT data. The likelihood when 2007 data, 2009 data, and
  the combined data set are used is shown in the three sets of
  panels. The parameter constraints derived from comparing the
    data to time-averaged images are overplotted in red contours. The
    parameter constraints are both offset from the constraints derived
    when considering the intrinsic variability of the model, and the
    allowed parameter range is incorrectly reduced compared to the
    true uncertainty reflected by the time-dependent analysis.
    }
\label{fig:models}
\end{figure*}

In this section, we apply the Bayesian method to early, limited EHT
visibility amplitude data. These EHT observations of Sgr~A* were
carried out in 2007 and 2009 with only three stations
\citep{2008Natur.455...78D, 2011ApJ...727L..36F}: the James Clerk
Maxwell Telescope in Hawaii (HI), the CARMA in
California (CA), and the SMT of the Arizona Radio Observatory (ARO) in Arizona (AZ). The
observations measured 18 visibility amplitudes in 2007 (HI-AZ: 4, AZ-CA:
14), and 51 (HI-AZ: 19, HI-CA: 12, AZ-CA: 20) in 2009.

In our comparison of the GRMHD models to the data, we only consider
values for the orientation $\xi$ from -90$^{\circ}$ to 90$^{\circ}$,
as before, because visibility amplitudes have a 180$^{\circ}$
degeneracy. Closure phase information is required to resolve the
ambiguity, and the closure phases were not measured well in these early
data.

In order to compare the simulations with real data, which are affected
by interstellar scattering, we incorporated the effects of scattering
using a simple prescription. Multiwavelength observations of Sgr~A*
show that its intrinsic size scales as ${\lambda}^2$ due to scattering
by electrons along the line of sight to the Galactic Center. The
scattering is described as an elliptical kernel
\citep{2006ApJ...648L.127B} having major and minor axes
\begin{equation}
\textrm{FWHM}_{\textrm{\small{major}}} = 1.309 (\lambda / 1 \ \textrm{cm})^2 \ \textrm{mas},
\end{equation}
and
\begin{equation}
\textrm{FWHM}_{\textrm{\small{minor}}} = 0.64 (\lambda / 1 \ \textrm{cm})^2 \ \textrm{mas}
\end{equation}
with a position angle of the major axis $\textrm{PA}=78^{\circ}$ east
of north. In the visibility space, this corresponds to an elliptical
taper of the visibility amplitude \citep{2014ApJ...795..134F}.
  Psaltis et al.\ (2015) revisited the inference of the scattering
  kernel parameters from observations and evaluated their
  uncertainties. \cite{2015ApJ...805..180J} described the refractive
scintillation in more detail, where the real effect of scattering is
likely more complicated than this simple convolution
prescription. For the purposes of our initial study, we will not
  consider further these two improvements on the modeling of the
  scattering kernel.

Figure~\ref{fig:models} shows the normalized posterior likelihoods
when comparing the EHT data to the five GRMHD models. Notably, the
observational data identify different physical parameters for the five
models, even though all models are physically plausible descriptions
of the structure of Sgr A*. For all five models, the 2009 EHT data
dominate the outcome because of the larger number of detections in
that campaign. Parameters with maximum likelihoods are given in
Table~\ref{table:modeleht} with 68\% errors. While there is some
agreement about the orientation parameter among most of the models
with limited data, it is clear that inferences about the spin
orientation are model specific and not general. Of course, the
  analysis presented here does not explore the effect of varying other
  parameters, such as the inclination.  Expanding the dimensionality of
  the parameter space for each model will most likely affect the
  inferred model parameters and their uncertainties.

Figure~\ref{fig:models} and Table~\ref{table:modeleht}
  also compare the likelihoods obtained when fitting time-dependent
  and time-averaged GRMHD images to the data. The time-averaged
analysis is similar to the time-resolved one, except that $P^{\rm
  sim}({\cal V})$ is composed of visibilities generated from a single
image obtained by averaging all the snapshots. As with the mock data,
 the lack of variability in the averaged model image results in a
  substantial underestimation of the uncertainties in the model
  parameters.

\begin{figure*}
\begin{center}
\includegraphics[width=16.5cm]{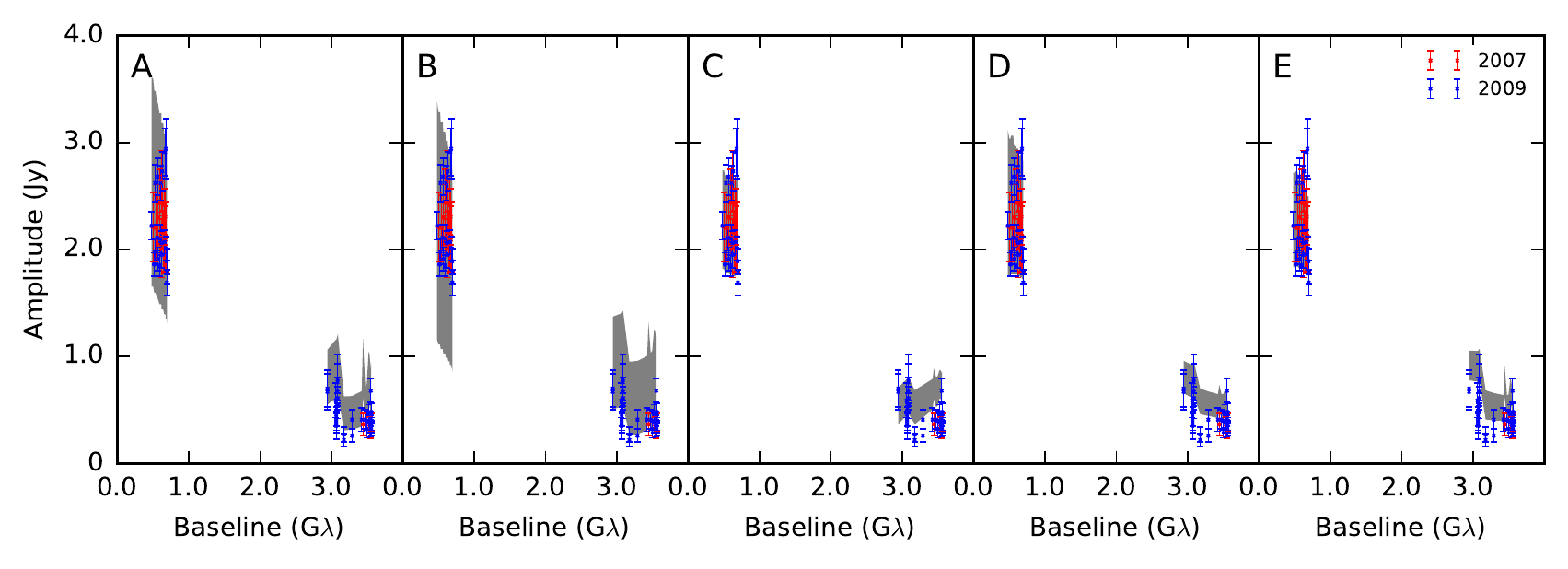} 
\end{center}
\caption{VLBI visibility amplitudes of Models A to E with the best-fit
  parameters listed in Table~\ref{table:modeleht} and EHT data. The
  visibility amplitudes are plotted with their $\pm 1 \sigma$ range
  in the shaded region. The EHT data include two days of
  observations in 2007, and three days observations in 2009. Although
  the simulated visibility amplitudes and observations appear to agree well with each
  other for all five models, marginal likelihoods and BIC show
  model E is the favored one with the limited EHT data.}
\label{fig:vis_bestfit}
\end{figure*}

Remarkably, comparing model evidences shows that the SANE
  Models A and B are decisively rejected in the time-dependent
  analysis, while they are among the most favored in the time-averaged
  analysis. This happens because, even though the images of Models A
  and B have an overall structure that matches the observations
  (compare, e.g., the data and the predictions of the five models at
  baselines larger than $\sim$ 3 G$\lambda$ in
  Figure~\ref{fig:vis_bestfit}), they are nevertheless disfavored in
  the time-dependent analysis because they show much larger
  variability than the current limited data (compare the range of
  visibilities predicted at $\lesssim$ 1 G$\lambda$ by all five
  models in Figure~\ref{fig:vis_bestfit}). Clearly, retaining knowledge of the variability predicted
  by a particular GRMHD (or other) simulation is important for
  evaluating how well that simulation matches the data.

The Bayesian evidence for the different models shown in
  Table~\ref{table:modeleht} seems to suggest that the existing EHT
  data favor MAD models over the SANE ones. However, it is important to
  emphasize here a requirement that needs to be satisfied before the
  Bayesian method can be applied to real data: that both the
  simulations and the observations have sampled adequately the
  variability at each baseline so that they cover the entire range of
  possibilities. This is clearly not the case for the early EHT data
  that we are using here, since they comprise only $\sim$10 hrs of
  total integration time.  As a result, while we present this analysis
  as a proof of principle, inferring model parameters from real EHT
  data based on our Bayesian method will be possible only after a
  substantial amount of data has been collected and their variability
  has been characterized.

There exist in the literature several analyses of the current EHT data
based on time-averaged models. \cite{2009ApJ...697...45B,
    2011ApJ...735..110B} modeled the accretion flow with a
time-independent, semi-analytic RIAF models and inferred the following as the most
probable black hole parameters: viewing angle\footnote{Note that
  \cite{2009ApJ...697...45B, 2011ApJ...735..110B} uses
  $\theta$ instead of $i$ for the observer's inclination angle.}
$i = {50^{\circ}}_{-10^{\circ}}^{+10^{\circ}}$, and position
  angle $\xi = {-20^{\circ}}_{-16^{\circ}}^{+31^{\circ}}$ for the 2007
  observations only, and $i = {68^{\circ}}_{-20^{\circ}}^{+5^{\circ}}$
  and $\xi = {-52^{\circ}}_{-15^{\circ}}^{+17^{\circ}}$ for the 2007
  and 2009 observations. \cite{2010ApJ...717.1092D, 2012JPhCS.372a2023D} performed
  three-dimensional, time-dependent GRMHD simulation, then averaged the simulated images to fit the 
  VLBI data. They provided
  $i = {50^{\circ}}_{-15^{\circ}}^{+35^{\circ}}$
  and $\xi = {-23^{\circ}}_{-22^{\circ}}^{+97^{\circ}}$ from the 2007 observation data, and
  $i = {60^{\circ}} \pm 15^{\circ}$
  and $\xi = {-70^{\circ}}_{-15^{\circ}}^{+86^{\circ}}$ using the 2007 and 2009 observations.\footnote{\cite{2010ApJ...717.1092D, 2012JPhCS.372a2023D} use 90\% confidence for the uncertainty of estimated parameters.}
  
Although these values are broadly consistent with each other and with
those we inferred here for most models, their uncertainties are
significantly tighter than what we found here for the SANE models
  (which are the only types of models considered earlier). This is not
  surprising because, as the above analysis suggests, fitting models
  that do not allow for source variability yields artificially small
  parameter uncertainties when the source is, in fact, variable.

\begin{deluxetable*}{ccccc|cccc}
\tablecolumns{9}
\tablewidth{0pc}
\tablecaption{Best-fit parameters and model evidences with EHT observations in 2007 and 2009}
\tablehead{\colhead{} & \multicolumn{4}{c}{Time-dependent analysis} & \multicolumn{4}{c}{Time-averaged analysis}\\ \colhead{} & \colhead{$\xi$} & \colhead{$F_{0}$} & \colhead{$\mathcal{L} (m\vert\textrm{data})$} & \colhead{$\triangle$BIC} & \colhead{$\xi$} & \colhead{$F_{0}$} & \colhead{$\mathcal{L} (m\vert\textrm{data})$} & \colhead{$\triangle$BIC}}
\startdata
Model A & ${-46^{\circ}}_{-6^{\circ}}^{+9^{\circ}}$ & $1.41_{-0.19}^{+0.23}$ & $1.84 \times 10^{-8}$ & 37.1 & ${-42^{\circ}}_{-3^{\circ}}^{+3^{\circ}}$ & $1.27_{-0.03}^{+0.03}$ & $5.98 \times 10^{-1}$ & 0.0\\
Model B & ${-36^{\circ}}_{-8^{\circ}}^{+43^{\circ}}$ & $1.23_{-0.22}^{+0.19}$ & $5.99 \times 10^{-15}$ & 68.2 & ${-46^{\circ}}_{-2^{\circ}}^{+2^{\circ}}$ & $1.37_{-0.03}^{+0.03}$ & $4.93 \times 10^{-2}$ & 4.5\\
Model C & ${50^{\circ}}_{-5^{\circ}}^{+6^{\circ}}$ & $1.11_{-0.06}^{+0.06}$ & $2.71 \times 10^{-2}$ & 6.9 & ${53^{\circ}}_{-5^{\circ}}^{+4^{\circ}}$ & $1.05_{-0.02}^{+0.03}$ & $3.62 \times 10^{-2}$ & 6.0\\
Model D & ${-37^{\circ}}_{-6^{\circ}}^{+14^{\circ}}$ & $2.44_{-0.15}^{+0.15}$ & $1.18 \times 10^{-3}$ & 15.7 & ${-32^{\circ}}_{-4^{\circ}}^{+11^{\circ}}$ & $2.24_{-0.05}^{+0.05}$ & $6.56 \times 10^{-4}$ & 16.0\\
Model E & ${-39^{\circ}}_{-3^{\circ}}^{+4^{\circ}}$ & $2.50_{-0.11}^{+0.12}$ & $1.00$ & 0.0 & ${-37^{\circ}}_{-3^{\circ}}^{+4^{\circ}}$ & $2.41_{-0.05}^{+0.05}$ & $1.00$ & 0.8
\enddata
\label{table:modeleht}
\end{deluxetable*}

\section{Conclusion}

In this work, we developed a statistical tool that will allow us to
interpret the variable EHT observational data in the context of
time-dependent model predictions. We tested the method for parameter
estimation and model selection statistics. We then applied this
  Bayesian statistical analysis to existing EHT observations from 2007
  and 2009, as a proof of principle, and investigated its applicability
  and statistical power.

We find that taking into account the time variability in the data
  and in the models increases the uncertainties in the inferred
  parameters beyond what is obtained when using time-averaged emission
  structures. Moreover, ignoring the variable nature of the data and of
  the simulations may lead to erroneous model selections.

In our analysis, we only considered two parameters -- the orientation
of the black hole spin vector in the sky plane and the flux normalization
-- in this paper. Additional parameters, such as the inclination angle
of the spin vector, the black hole spin and the plasma parameters of
the accretion flow, can also be investigated with this method by
expanding the parameter space and injecting corresponding simulation
data sets.

Finally, the method presented in this paper is quite general and can
work with any time-resolved simulation. Although only visibility
amplitudes were included in our testing, other observables such as
closure phase and amplitude can easily be incorporated with future
observations with additional baselines.

\acknowledgments
J.K. and D.P.M. acknowledge support from NSF grants AST-1207752 and AST-1440254.
C.K.C., F.O., and D.P. were partially supported by NASA/NSF TCAN award
NNX14AB48G and NSF grants AST~1108753 and AST~1312034.
L.M. acknowledges support from NFS GRFP grant DGE~1144085.
All ray-tracing calculations were performed with the \texttt{El~Gato}
GPU cluster at the University of Arizona that is funded by NSF award
1228509.

\bibliographystyle{apj}
\bibliography{ms.bib} 

\begin{thebibliography}{}
\expandafter\ifx\csname natexlab\endcsname\relax\def\natexlab#1{#1}\fi

\bibitem[{Bardeen(1973)}]{1973blho.conf..215B}
Bardeen, J.~M. 1973, in Black Holes (Les Astres Occlus), ed. C.~D. {\&}. B. S.
  D. N.~Y. Gordon \& Breach), 215--239

\bibitem[{Bartko {et~al.}(2009)Bartko, Martins, Fritz, Genzel, Levin, Perets,
  Paumard, Nayakshin, Gerhard, Alexander, Dodds-Eden, Eisenhauer, Gillessen,
  Mascetti, Ott, Perrin, Pfuhl, Reid, Rouan, Sternberg, \&
  Trippe}]{2009ApJ...697.1741B}
Bartko, H., Martins, F., Fritz, T.~K., {et~al.} 2009, The Astrophysical
  Journal, 697, 1741

\bibitem[{Bower {et~al.}(2006)Bower, Goss, Falcke, Backer, \&
  Lithwick}]{2006ApJ...648L.127B}
Bower, G.~C., Goss, W.~M., Falcke, H., Backer, D.~C., \& Lithwick, Y. 2006, The
  Astrophysical Journal, 648, L127

\bibitem[{Bower {et~al.}(2015)Bower, Markoff, Dexter, Gurwell, Moran,
  Brunthaler, Falcke, Fragile, Maitra, Marrone, Peck, Rushton, \&
  Wright}]{2015ApJ...802...69B}
Bower, G.~C., Markoff, S., Dexter, J., {et~al.} 2015, The Astrophysical
  Journal, 802, 69

\bibitem[{Broderick {et~al.}(2009)Broderick, Fish, Doeleman, \&
  Loeb}]{2009ApJ...697...45B}
Broderick, A.~E., Fish, V.~L., Doeleman, S.~S., \& Loeb, A. 2009, The
  Astrophysical Journal, 697, 45

\bibitem[{Broderick {et~al.}(2011)Broderick, Fish, Doeleman, \&
  Loeb}]{2011ApJ...735..110B}
---. 2011, The Astrophysical Journal, 735, 110

\bibitem[{Chan {et~al.}(2009)Chan, Liu, Fryer, Psaltis, {\"O}zel, Rockefeller,
  \& Melia}]{2009ApJ...701..521C}
Chan, C.-K., Liu, S., Fryer, C.~L., {et~al.} 2009, The Astrophysical Journal,
  701, 521

\bibitem[{Chan {et~al.}(2013)Chan, Psaltis, \& {\"O}zel}]{2013ApJ...777...13C}
Chan, C.-K., Psaltis, D., \& {\"O}zel, F. 2013, The Astrophysical Journal, 777,
  13

\bibitem[{Chan {et~al.}(2015{\natexlab{a}})Chan, Psaltis, {\"O}zel, Medeiros,
  Marrone, Sa{\c{d}}owski, \& Narayan}]{2015ApJ...812..103C}
Chan, C.-K., Psaltis, D., {\"O}zel, F., {et~al.} 2015{\natexlab{a}}, The
  Astrophysical Journal, 812, 103

\bibitem[{Chan {et~al.}(2015{\natexlab{b}})Chan, Psaltis, {\"O}zel, Narayan, \&
  Sa{\c{d}}owski}]{2015ApJ...799....1C}
Chan, C.-K., Psaltis, D., {\"O}zel, F., Narayan, R., \& Sa{\c{d}}owski, A.
  2015{\natexlab{b}}, The Astrophysical Journal, 799, 1

\bibitem[{Chatzopoulos {et~al.}(2015)Chatzopoulos, Fritz, Gerhard, Gillessen,
  Wegg, Genzel, \& Pfuhl}]{2015MNRAS.447..948C}
Chatzopoulos, S., Fritz, T.~K., Gerhard, O., {et~al.} 2015, Monthly Notices of
  the Royal Astronomical Society, 447, 948

\bibitem[{Dexter {et~al.}(2010)Dexter, Agol, Fragile, \&
  McKinney}]{2010ApJ...717.1092D}
Dexter, J., Agol, E., Fragile, P.~C., \& McKinney, J.~C. 2010, The
  Astrophysical Journal, 717, 1092

\bibitem[{Dexter {et~al.}(2012)Dexter, Agol, Fragile, \&
  McKinney}]{2012JPhCS.372a2023D}
---. 2012, Journal of Physics: Conference Series, 372, 012023

\bibitem[{Dexter {et~al.}(2014)Dexter, Kelly, Bower, Marrone, Stone, \&
  Plambeck}]{2014MNRAS.442.2797D}
Dexter, J., Kelly, B., Bower, G.~C., {et~al.} 2014, Monthly Notices of the
  Royal Astronomical Society, 442, 2797

\bibitem[{Doeleman {et~al.}(2001)Doeleman, Shen, Rogers, Bower, Wright, Zhao,
  Backer, Crowley, Freund, Ho, Lo, \& Woody}]{2001AJ....121.2610D}
Doeleman, S.~S., Shen, Z.~Q., Rogers, A. E.~E., {et~al.} 2001, The Astronomical
  Journal, 121, 2610

\bibitem[{Doeleman {et~al.}(2008)Doeleman, Weintroub, Rogers, Plambeck, Freund,
  Tilanus, Friberg, Ziurys, Moran, Corey, Young, Smythe, Titus, Marrone,
  Cappallo, Bock, Bower, Chamberlin, Davis, Krichbaum, Lamb, Maness, Niell,
  Roy, Strittmatter, Werthimer, Whitney, \& Woody}]{2008Natur.455...78D}
Doeleman, S.~S., Weintroub, J., Rogers, A. E.~E., {et~al.} 2008, Nature, 455,
  78

\bibitem[{Doeleman {et~al.}(2012)Doeleman, Fish, Schenck, Beaudoin, Blundell,
  Bower, Broderick, Chamberlin, Freund, Friberg, Gurwell, Ho, Honma, Inoue,
  Krichbaum, Lamb, Loeb, Lonsdale, Marrone, Moran, Oyama, Plambeck, Primiani,
  Rogers, Smythe, SooHoo, Strittmatter, Tilanus, Titus, Weintroub, Wright,
  Young, \& Ziurys}]{2012Sci...338..355D}
Doeleman, S.~S., Fish, V.~L., Schenck, D.~E., {et~al.} 2012, Science, 338, 355

\bibitem[{Falcke \& Markoff(2013)}]{2013CQGra..30x4003F}
Falcke, H., \& Markoff, S.~B. 2013, Classical and Quantum Gravity, 30, 244003

\bibitem[{Falcke {et~al.}(2000)Falcke, Melia, \& Agol}]{2000ApJ...528L..13F}
Falcke, H., Melia, F., \& Agol, E. 2000, The Astrophysical Journal, 528, L13

\bibitem[{Fish {et~al.}(2011)Fish, Doeleman, Beaudoin, Blundell, Bolin, Bower,
  Chamberlin, Freund, Friberg, Gurwell, Honma, Inoue, Krichbaum, Lamb, Marrone,
  Moran, Oyama, Plambeck, Primiani, Rogers, Smythe, SooHoo, Strittmatter,
  Tilanus, Titus, Weintroub, Wright, Woody, Young, \&
  Ziurys}]{2011ApJ...727L..36F}
Fish, V.~L., Doeleman, S.~S., Beaudoin, C., {et~al.} 2011, The Astrophysical
  Journal Letters, 727, L36

\bibitem[{Fish {et~al.}(2014)Fish, Johnson, Lu, Doeleman, Bouman, Zoran,
  Freeman, Psaltis, Narayan, Pankratius, Broderick, Gwinn, \&
  Vertatschitsch}]{2014ApJ...795..134F}
Fish, V.~L., Johnson, M.~D., Lu, R.-S., {et~al.} 2014, The Astrophysical
  Journal, 795, 134

\bibitem[{Gammie {et~al.}(2003)Gammie, McKinney, \&
  T{\'o}th}]{2003ApJ...589..444G}
Gammie, C.~F., McKinney, J.~C., \& T{\'o}th, G. 2003, The Astrophysical
  Journal, 589, 444

\bibitem[{Ghez {et~al.}(2008)Ghez, Salim, Weinberg, Lu, Do, Dunn, Matthews,
  Morris, Yelda, Becklin, Kremenek, Milosavljevic, \&
  Naiman}]{2008ApJ...689.1044G}
Ghez, A.~M., Salim, S., Weinberg, N.~N., {et~al.} 2008, The Astrophysical
  Journal, 689, 1044

\bibitem[{Gillessen {et~al.}(2009{\natexlab{a}})Gillessen, Eisenhauer, Fritz,
  Bartko, Dodds-Eden, Pfuhl, Ott, \& Genzel}]{2009ApJ...707L.114G}
Gillessen, S., Eisenhauer, F., Fritz, T.~K., {et~al.} 2009{\natexlab{a}}, The
  Astrophysical Journal Letters, 707, L114

\bibitem[{Gillessen {et~al.}(2009{\natexlab{b}})Gillessen, Eisenhauer, Trippe,
  Alexander, Genzel, Martins, \& Ott}]{2009ApJ...692.1075G}
Gillessen, S., Eisenhauer, F., Trippe, S., {et~al.} 2009{\natexlab{b}}, The
  Astrophysical Journal, 692, 1075

\bibitem[{Gold {et~al.}(2016)Gold, McKinney, Johnson, \&
  Doeleman}]{Gold:2016ud}
Gold, R., McKinney, J.~C., Johnson, M.~D., \& Doeleman, S.~S. 2016,
  arXiv:1601.05550

\bibitem[{Goldston {et~al.}(2005)Goldston, Quataert, \&
  Igumenshchev}]{2005ApJ...621..785G}
Goldston, J.~E., Quataert, E., \& Igumenshchev, I.~V. 2005, The Astrophysical
  Journal, 621, 785

\bibitem[{Huang {et~al.}(2007)Huang, Cai, Shen, \& Yuan}]{2007MNRAS.379..833H}
Huang, L., Cai, M., Shen, Z.-Q., \& Yuan, F. 2007, Monthly Notices of the Royal
  Astronomical Society, 379, 833

\bibitem[{Johannsen \& Psaltis(2010)}]{2010ApJ...718..446J}
Johannsen, T., \& Psaltis, D. 2010, The Astrophysical Journal, 718, 446

\bibitem[{Johannsen {et~al.}(2012)Johannsen, Psaltis, Gillessen, Marrone,
  {\"O}zel, Doeleman, \& Fish}]{2012ApJ...758...30J}
Johannsen, T., Psaltis, D., Gillessen, S., {et~al.} 2012, The Astrophysical
  Journal, 758, 30

\bibitem[{Johnson \& Gwinn(2015)}]{2015ApJ...805..180J}
Johnson, M.~D., \& Gwinn, C.~R. 2015, The Astrophysical Journal, 805, 180

\bibitem[{Kass \& Raftery(1995)}]{Kass:1995eh}
Kass, R.~E., \& Raftery, A.~E. 1995, Journal of the American Statistical
  Association, 90, 773

\bibitem[{Liddle(2007)}]{2007MNRAS.377L..74L}
Liddle, A.~R. 2007, Monthly Notices of the Royal Astronomical Society: Letters,
  377, L74

\bibitem[{Luminet(1979)}]{1979A&A....75..228L}
Luminet, J.~P. 1979, Astronomy and Astrophysics, 75, 228

\bibitem[{Marrone {et~al.}(2006)Marrone, Moran, Zhao, \&
  Rao}]{2006JPhCS..54..354M}
Marrone, D.~P., Moran, J.~M., Zhao, J.-H., \& Rao, R. 2006, Journal of Physics:
  Conference Series, 54, 354

\bibitem[{Marrone {et~al.}(2008)Marrone, Baganoff, Morris, Moran, Ghez,
  Hornstein, Dowell, Mu{\~n}oz, Bautz, Ricker, Brandt, Garmire, Lu, Matthews,
  Zhao, Rao, \& Bower}]{2008ApJ...682..373M}
Marrone, D.~P., Baganoff, F.~K., Morris, M.~R., {et~al.} 2008, The
  Astrophysical Journal, 682, 373

\bibitem[{McKinney(2006)}]{2006MNRAS.368.1561M}
McKinney, J.~C. 2006, Monthly Notices of the Royal Astronomical Society, 368,
  1561

\bibitem[{McKinney \& Blandford(2009)}]{2009MNRAS.394L.126M}
McKinney, J.~C., \& Blandford, R.~D. 2009, Monthly Notices of the Royal
  Astronomical Society: Letters, 394, L126

\bibitem[{Medeiros {et~al.}(2016)Medeiros, Chan, {\"O}zel, Psaltis, Kim,
  Marrone, \& Sa{\c{d}}owski}]{2016arXiv160106799M}
Medeiros, L., Chan, C.-K., {\"O}zel, F., {et~al.} 2016, arXiv:1601.06799

\bibitem[{Mo{\'{s}}cibrodzka \& Falcke(2013)}]{2013A&A...559L...3M}
Mo{\'{s}}cibrodzka, M., \& Falcke, H. 2013, Astronomy and Astrophysics, 559, L3

\bibitem[{Mo{\'{s}}cibrodzka {et~al.}(2009)Mo{\'{s}}cibrodzka, Gammie, Dolence,
  Shiokawa, \& Leung}]{2009ApJ...706..497M}
Mo{\'{s}}cibrodzka, M., Gammie, C.~F., Dolence, J.~C., Shiokawa, H., \& Leung,
  P.~K. 2009, The Astrophysical Journal, 706, 497

\bibitem[{Narayan {et~al.}(1998)Narayan, Mahadevan, Grindlay, Popham, \&
  Gammie}]{1998ApJ...492..554N}
Narayan, R., Mahadevan, R., Grindlay, J.~E., Popham, R.~G., \& Gammie, C. 1998,
  The Astrophysical Journal, 492, 554

\bibitem[{Narayan {et~al.}(2012)Narayan, Sa{\c{d}}owski, Penna, \&
  Kulkarni}]{2012MNRAS.426.3241N}
Narayan, R., Sa{\c{d}}owski, A., Penna, R.~F., \& Kulkarni, A.~K. 2012, Monthly
  Notices of the Royal Astronomical Society, 426, 3241

\bibitem[{{\"O}zel {et~al.}(2000){\"O}zel, Psaltis, \&
  Narayan}]{2000ApJ...541..234O}
{\"O}zel, F., Psaltis, D., \& Narayan, R. 2000, The Astrophysical Journal, 541,
  234

\bibitem[{Psaltis {et~al.}(2015)Psaltis, Narayan, Fish, Broderick, Loeb, \&
  Doeleman}]{2015ApJ...798...15P}
Psaltis, D., Narayan, R., Fish, V.~L., {et~al.} 2015, The Astrophysical
  Journal, 798, 15

\bibitem[{Sa{\c{d}}owski {et~al.}(2013)Sa{\c{d}}owski, Narayan, Penna, \&
  Zhu}]{2013MNRAS.436.3856S}
Sa{\c{d}}owski, A., Narayan, R., Penna, R., \& Zhu, Y. 2013, Monthly Notices of
  the Royal Astronomical Society, 436, 3856

\bibitem[{Shcherbakov {et~al.}(2012)Shcherbakov, Penna, \&
  McKinney}]{2012ApJ...755..133S}
Shcherbakov, R.~V., Penna, R.~F., \& McKinney, J.~C. 2012, The Astrophysical
  Journal, 755, 133

\bibitem[{{Thompson} {et~al.}(2001){Thompson}, {Moran}, \& {Swenson}}]{TMS2001}
{Thompson}, A.~R., {Moran}, J.~M., \& {Swenson}, Jr., G.~W. 2001,
  Interferometry and Synthesis in Radio Astronomy (2nd ed.; New York: Wiley)

\bibitem[{Yuan \& Narayan(2014)}]{2014ARA&A..52..529Y}
Yuan, F., \& Narayan, R. 2014, Annual Review of Astronomy and Astrophysics, 52,
  529

\bibitem[{Yuan {et~al.}(2003)Yuan, Quataert, \& Narayan}]{2003ApJ...598..301Y}
Yuan, F., Quataert, E., \& Narayan, R. 2003, The Astrophysical Journal, 598,
  301

\bibitem[{Yusef-Zadeh {et~al.}(2009)Yusef-Zadeh, Bushouse, Wardle, Heinke,
  Roberts, Dowell, Brunthaler, Reid, Martin, Marrone, Porquet, Grosso,
  Dodds-Eden, Bower, Wiesemeyer, Miyazaki, Pal, Gillessen, Goldwurm, Trap, \&
  Maness}]{2009ApJ...706..348Y}
Yusef-Zadeh, F., Bushouse, H., Wardle, M., {et~al.} 2009, The Astrophysical
  Journal, 706, 348

\end{thebibliography}
\end{document}